\documentclass[16pt,a4]{article}
\usepackage{epsfig}
\usepackage{setspace}
\title{Studying Three Phase Supply in School}
\author{Amit Kumar Singhal \& P.Arun\\
Department of Physics and Electronics,\\
S.G.T.B. Khalsa College\\
University of Delhi, Delhi 110 007, India.}

\begin{document}
\maketitle
\begin{abstract}
The power distribution of nearly all major countries have accepted 3-phase
distribution as a standard. With increasing power requirements of
instrumentation today even a small physics laboratory requires 3-phase supply.
While physics students are given an introduction of this 
in passing, no experiment work is done with 3-phase supply 
due to the sheer possibility of accidents while working with
such large powers. We believe a conceptual understanding of 3-phase supply
would be useful for physics students with hands on experience using a 
simple circuit that can be assembled even in a high school laboratorys. 
\end{abstract}

\section*{Introduction}

Edison's invention of direct current (DC) preceeded Tesla's invention of
alternating current (AC). However, once there were two possible modes of 
power available, an inevitable debate on the merits and demerits of the two 
started that led to
what is called the ``war of currents''. The inventor's became adversaries 
with Edison promoting direct current (DC) for electric power
distribution over the alternating current (AC). The ``war of currents'' got
so bitter that both inventors lost a lot of money and rumors have it, 
their Nobel prize \cite{r1}.

Today the debate is more or less resolved with AC being accepted as the best
method for electric power distribution, especially where the power
requirement is large. It can be appreciated that since direct current can not 
be trivially stepped up or stepped down, the same voltage level is 
transmitted as required by the load. This resulted in large 
transmission losses.

Transmission loss takes place due to heat dissipation along the current
carrying wires used for delivering power from generation point to consumer. 
The initial transmission networks laid were of copper, which is one of the 
best conductors with low resistivity. Even with low resitivity, since the
length of the transmission wires involved are large, they offered finite 
and non-negligiable resitance. Thus introducing power loss during 
transmission. Mathematically power dissipated is gien as
\begin{eqnarray}
P_{dis}=I^2R=\rho \left({l \over A}\right)I^2
\end{eqnarray}
where I, ${\rm \rho}$, l and A are the rms (root mean square) 
current, wire's resistivity, it's length and cross-sectional area
respectively.  
Thus, for transmitting a given power with minimum power loss, one would have
to reduce the current while increasing the voltage. This is exactly what a
transformer does for AC. Thus for DC, transmission loss can only be 
minimised by using thicker copper
wires. In turn, only fairly low DC power can be transmitted. Since, AC power
can be stepped-up and down easily using transformers, the issue of
transmission loss over thinner wires can be economically addressed.

With industrialization and increasing demands for higher levels of power,
even a AC distribution is not enough. Hence, today distribution
has moved to polyphase (m) distribution. Polyphase voltages are also AC
voltages but made up of multiple sinosudial varying voltages. 
Of the possible polyphases, the three-phase supply
is the most popular. It refers to three voltages that differ in phase 
by ${\rm
2\pi/m=2\pi/3=120^o}$ degrees from each other. The voltages go through their 
maxima in a regular order, after every ${\rm 120^o}$. The phase sequence are 
named `A', `B' and `C'. 

\begin{figure}[t!!]
\begin{center}
\includegraphics[width=3.25in,angle=-0]{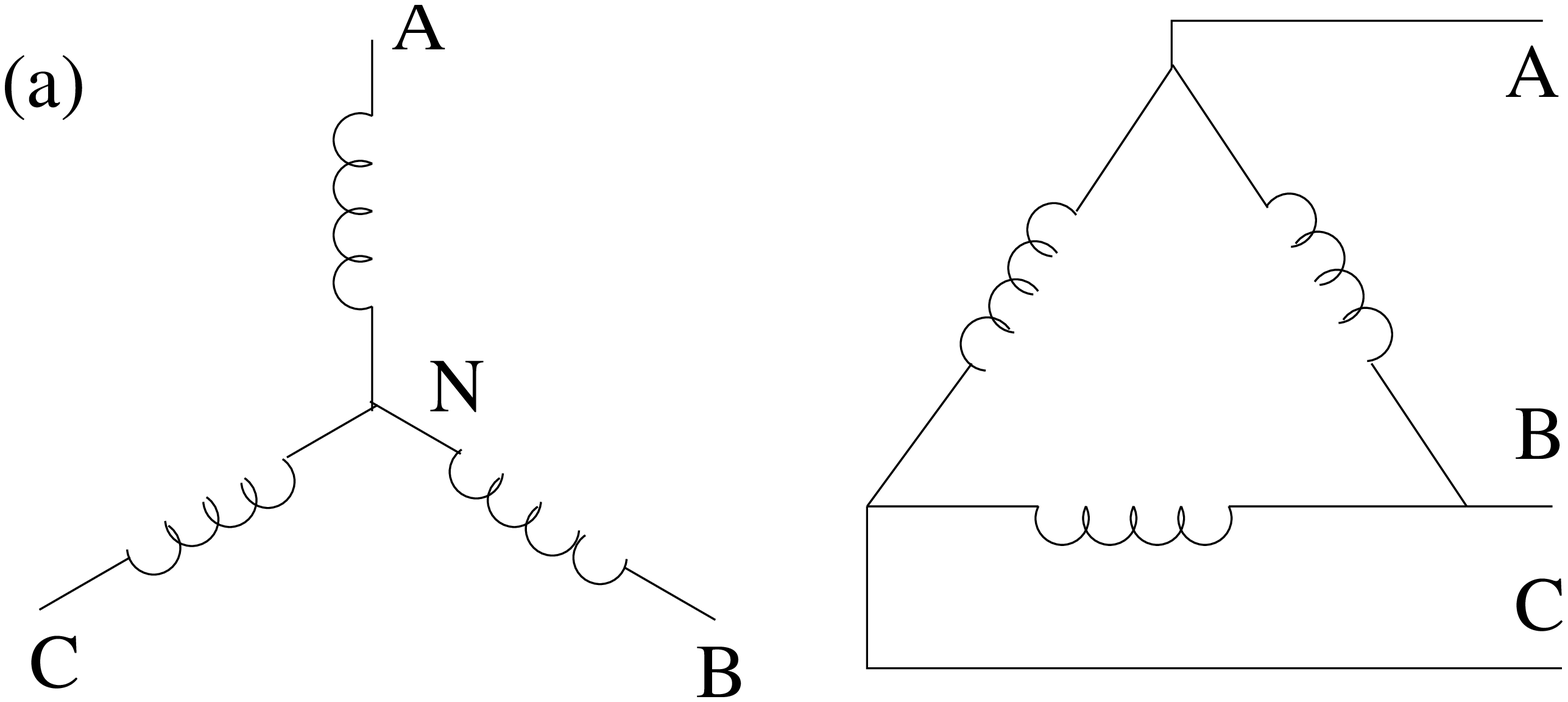}
\vskip 0.5cm
\includegraphics[width=2.25in,angle=-0]{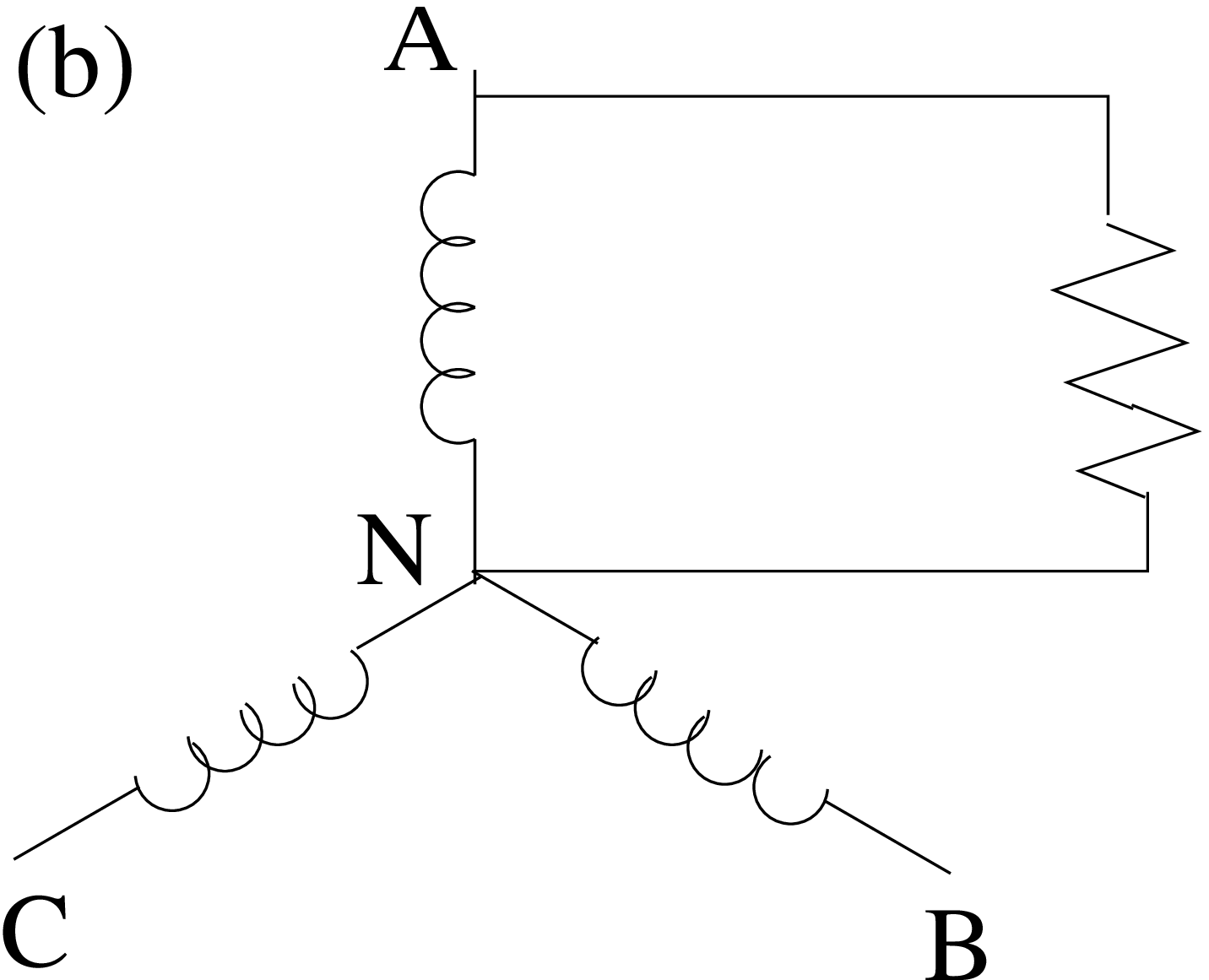}
\vskip 0.5cm
\includegraphics[width=2.25in,angle=-0]{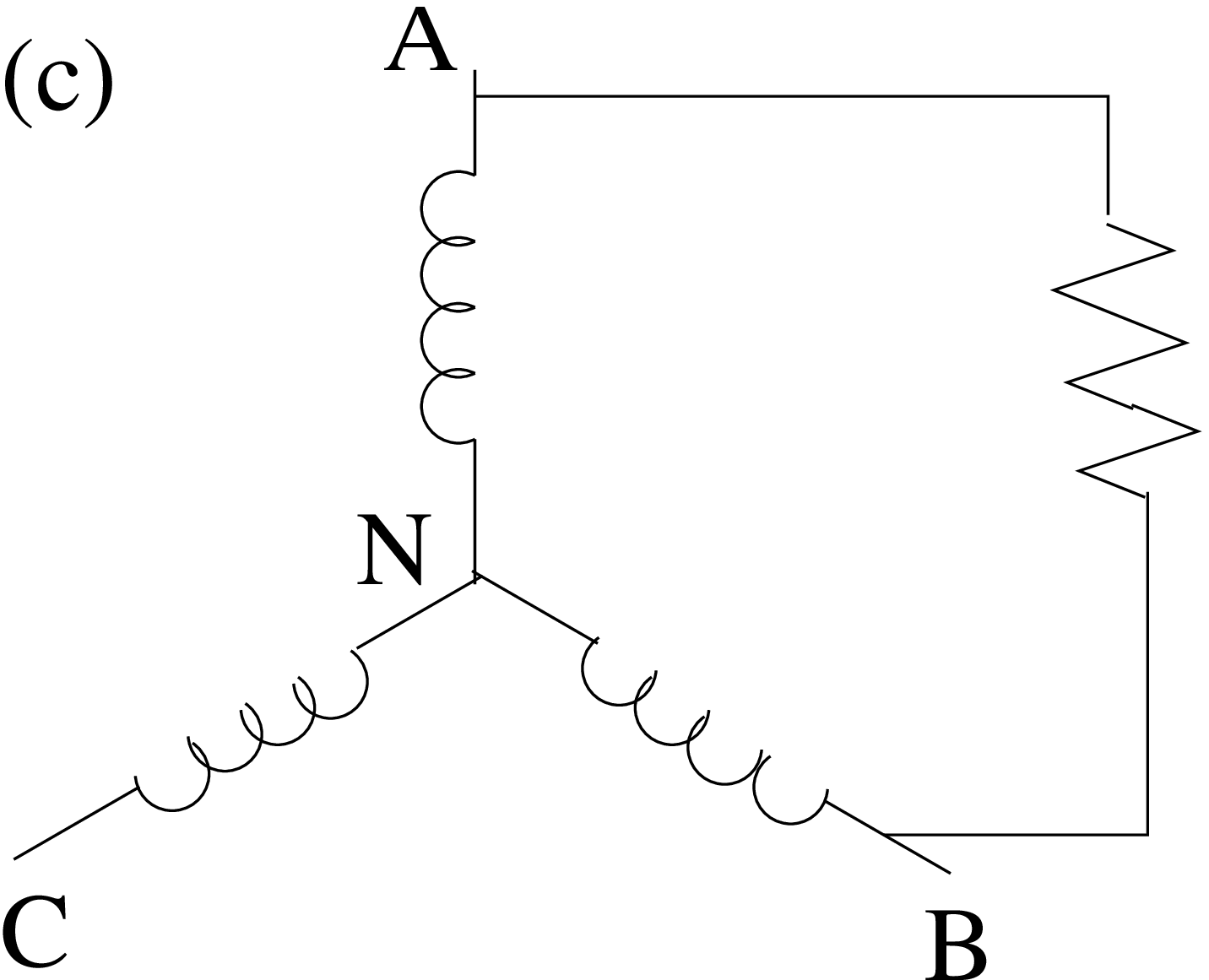}
\caption{(a) shows the two possible configurations of transformers used in 
3-phase supply. (b) and (c) shows examples of how load can be connected to
the star transformer. }
\end{center}
\end{figure}

The three phases are generally distributed using three 
wires. The phases are separated or collected at the load side using special
transformers, called ``star'' or ``delta'' transformers (fig~1a). Each wire
has the same current carrying capability as in case of single phase supply. 
However, the power delivered at the load can be far greater depending on how
the two potential levels are selected. Consider, in the star transformer the
load is connected between point `A' and 'neutral' (fig~1b) the output 
waveform would be that of phase A, with the neutral point acting as 
`zero potential' point. The power delivered to the load is ${\rm
V_m^2/2R_L}$. However, if the load is connected between point `A' and `B',
the net potential difference can be found (first in general) from
\begin{figure}[t!!]
\includegraphics[width=4in,angle=-90]{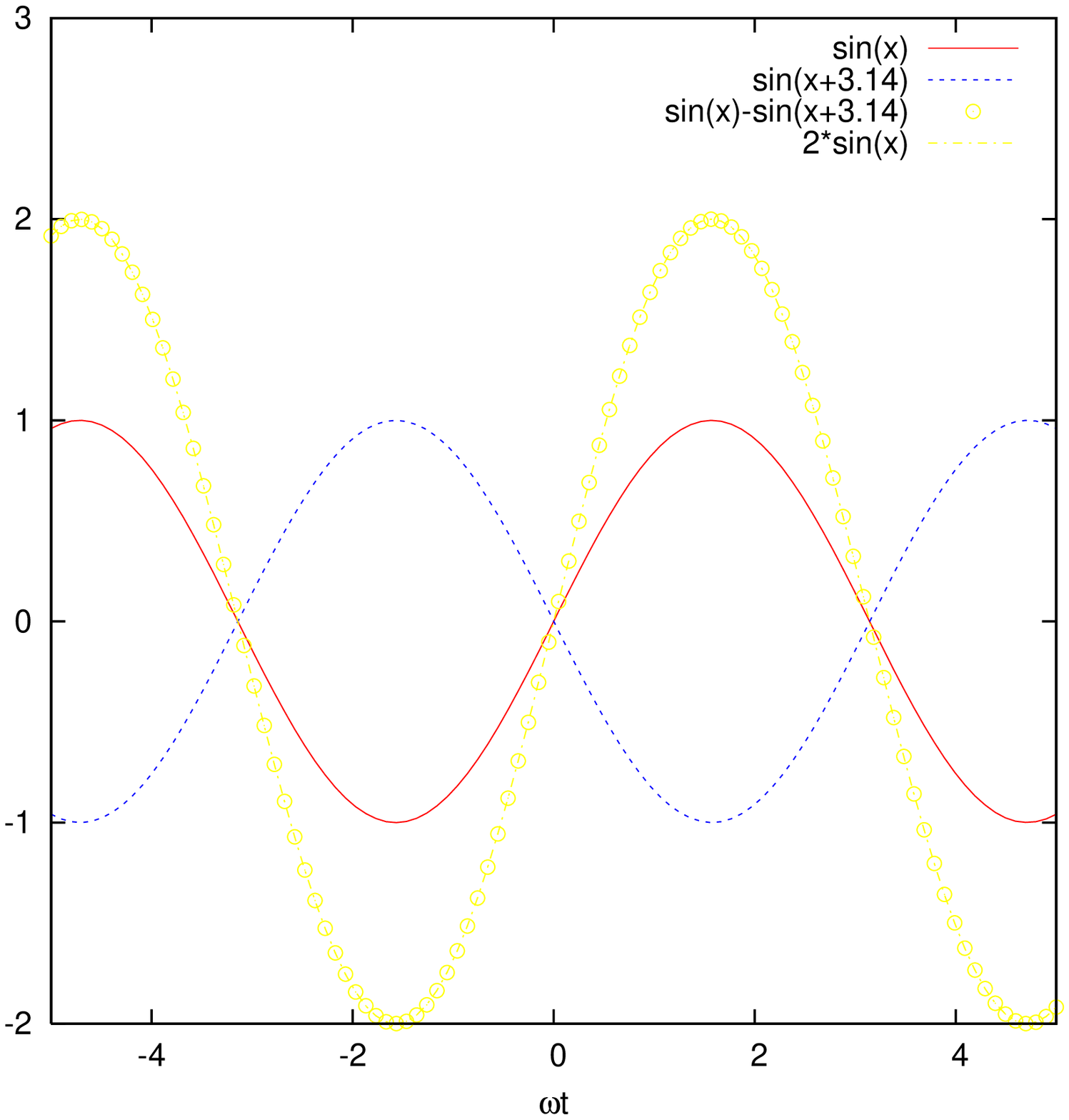}
\caption{The 3 phase supply carrying voltage of peak value, ${\rm V_m}$ can
be used to generate voltage of peak value ${\rm 2V_m}$.}   
\end{figure}

\begin{eqnarray}
V_{AB} &=&V_msin(\theta)-V_msin(\theta+\phi)\nonumber\\
&=&-2V_mcos\left({2\theta+\phi \over 2}\right)sin\left({\phi \over 
2}\right)\nonumber
\end{eqnarray} 
where ${\rm V_m}$ is the peak voltage and ${\rm \phi}$ the phase difference
between the two phases.
For a 2-phase supply, where the phase difference between the two waveforms
is ${\rm \pi}$, the potential difference would be
\begin{eqnarray}
V_{AB} &=&-2V_mcos\left(\theta+{\pi \over 2}\right)sin\left({\pi \over 
2}\right)\nonumber\\
&=&2V_msin(\theta)\nonumber
\end{eqnarray} 
This mathematics is graphically represented in fig~2. It can be appreciated 
that the power delivered to the load now would be ${\rm 2V_m^2/R_L}$.
This example shows how polyphase transmissions delivers more power.

While these are important issues to be appreciated by a physicist setting 
up a laboratory with instruments that require high power, 
schools and colleges do not
help develop this concept with hand-on experimentation for fear of possible
accidents that could even prove fatal. While kits are present in engineering
colleges for electrical engineering, not much is found in the literature in
way of simple experiemnts for physics and electronics students. In this
direction, we have designed and tested a simple circuit, 
where two phases are generated from a
sinusodial wave taken from a function generator.

\begin{figure}[t!!]
\begin{center}
\includegraphics[width=4.5in,angle=-0]{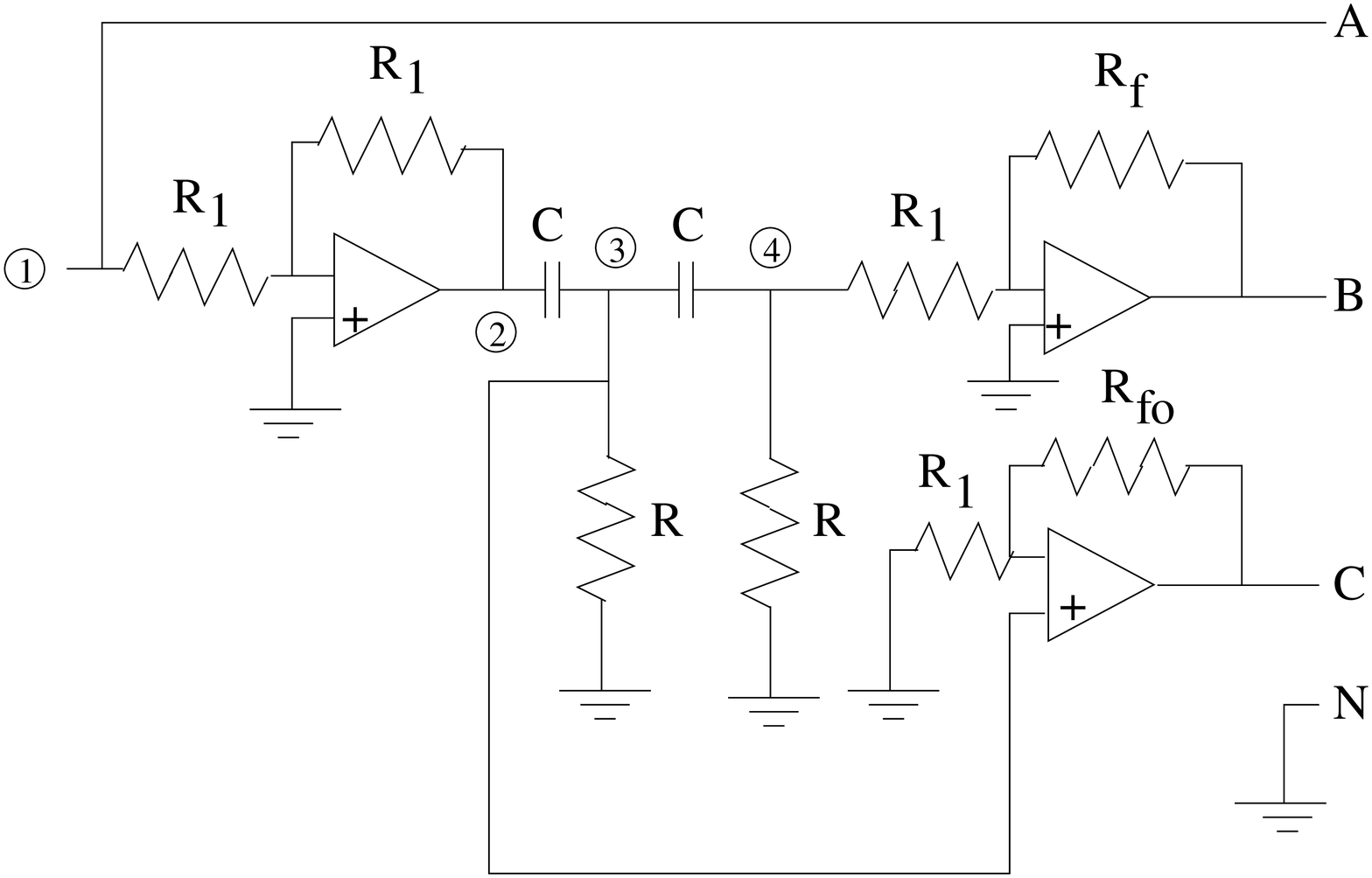}
\caption{Simple circuit for simulating the 3-phase supply used in modern
power distribution.}
\end{center}
\end{figure}

\section*{The Circuit}
\begin{figure}[h!!]
\begin{center}
\includegraphics[width=5in,angle=-0]{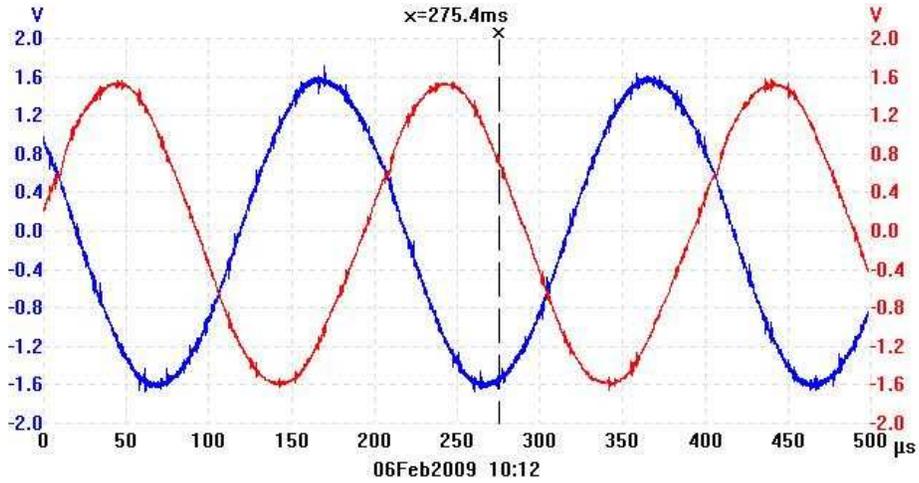}
\includegraphics[width=1.65in,angle=-90]{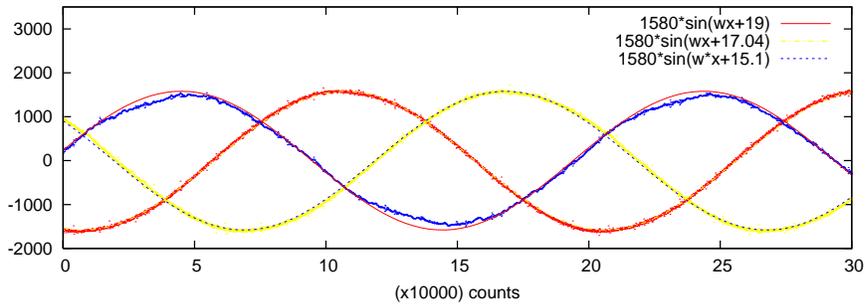}
\caption{The two Phases (A and C) as seen by the dual trace picoscope 2202.
Also, shown is the generated 3-phase supply using sinusodial wave from 
function generator (Phase A) and Phase B and Phase C using phase shift 
circuits. }
\end{center}
\end{figure}

The proposed circuit requires three opamps, of which the first opamp is
assembled in an inverting configuration \cite{opamp}. The sinusodial input
from a function generator (shown in fig~3 as `1') is given as an input to
this inverting amplifier. This signal also acts as one of the 3 phases,
namely phase A. The gain of this amplifier is kept as unity with
${\rm R_1=100K\Omega}$. This circuit acts as a buffer preventing any 
loading by successive circuits and introduces a phase change of ${\rm \pi}$
between output and input waveforms 
(as indicative of the name ``inverting'' amplifier).

The RC circuit is designed to introduce a phase difference of ${\rm 120^o}$
for the selected frequency. However, a single combination of RC can
at the most introduce a phase difference equal to or less than ${\rm 90^o}$.
Hence, two sections are used, with each RC section introducing 
a phase shift of ${\rm 60^o}$. The values of R and C are selected using the
formula \cite{arun}
\begin{eqnarray}
RC={1 \over \omega tan \phi}
\end{eqnarray}

{\sl {In our study, for an 5KHz wave provided from the function 
generator, we
selected ${\rm R=18K\Omega}$ and ${\rm C=0.001\mu F}$.}}
Hence points `3' and `4' of the circuit would be ${\rm 240^o\,(=180^o+60^o)}$
and ${\rm 300^o\,(=240^o+60^o)}$ respectively out of phase with respect to
the input. The impedence networks act as potential dividers and hence the
voltage levels at `3' and `4' would be lower than that given as input. Since
the three phases would have the same amplitude, the signals at `3' and `4'
would have to be amplified. We use as inverting amplifier again for point
`4'. The voltage gain of this amplifier is give by ${\rm R_f/R_1}$. Hence,
${\rm R_f}$ has to be varied till the voltage signals amplitude is identical
to the input signal. Also, this inverting amplifier introduces a phase
change of ${\rm \pi}$. The net phase difference hence would be ${\rm 480^o\,
(=300^o+180^o)}$ which is nothing but a phase difference of ${\rm 120^o\,
(=2\pi^o+120^o)}$. The output of this op amp circuit would be phase `B'.

Similar amplitude correction is required for the wave collected at point
`3'. However, here we select a non-inverting amplifier circuit whose gain is
given as ${\rm 1+R_{fo}/R_1}$ and as the name suggests does not introduce
any further phase shifts {\sl {(${\rm R_f}$ and ${\rm R_{fo}}$ used were 
${\rm 1M\Omega}$ pots and typically had values greater than 
${\rm 100K\Omega}$)}}. Hence, the net shift at the output remains the
same as that at point `3' at ${\rm 240^o}$ or ${\rm 240^o\,(=2\pi-120^o)}$.
This is phase C of the simulated 3-phase supply. 
\begin{figure}[h!!]
\begin{center}
\includegraphics[width=2in,angle=-0]{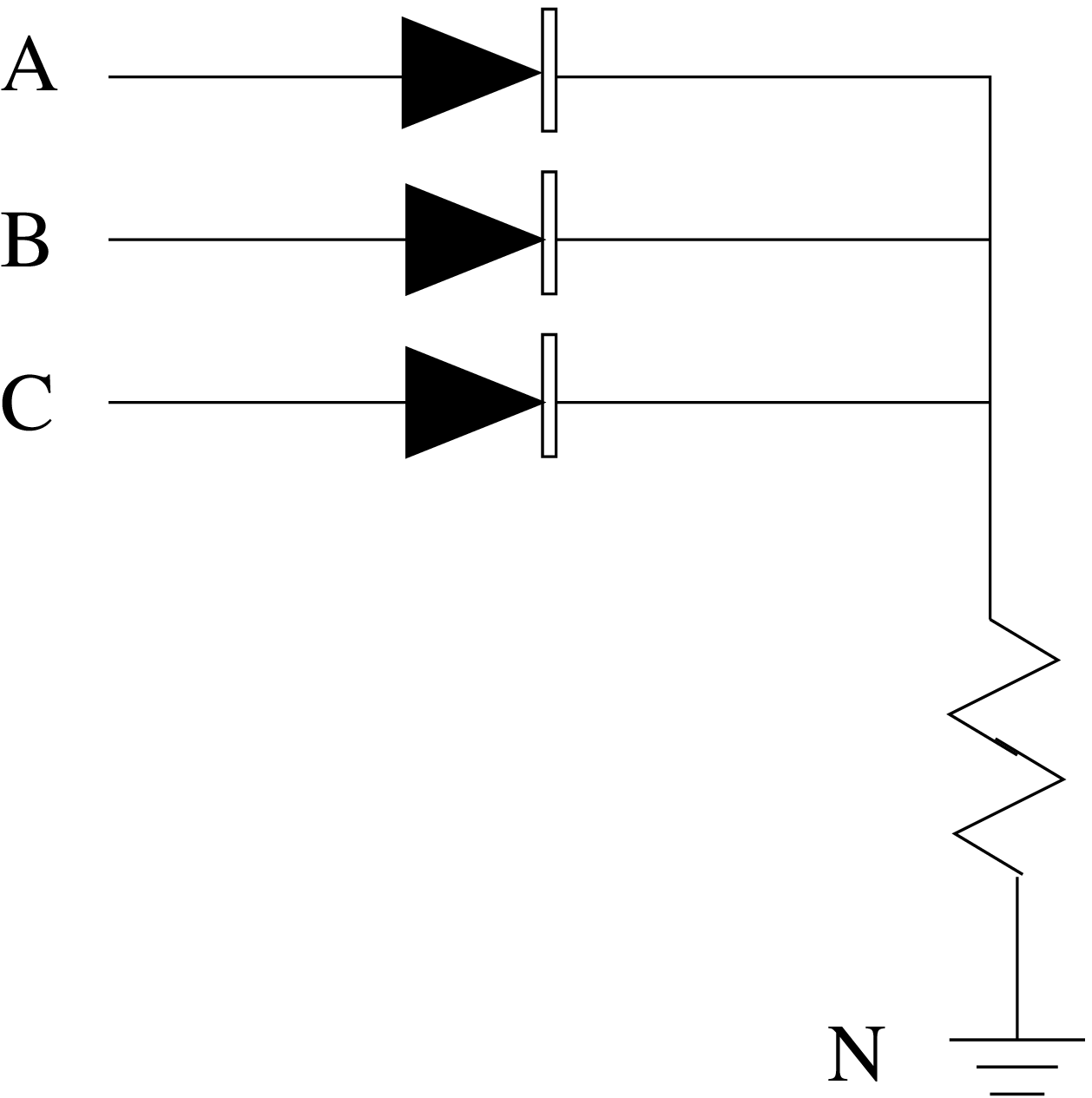}
\includegraphics[width=4.5in,angle=-0]{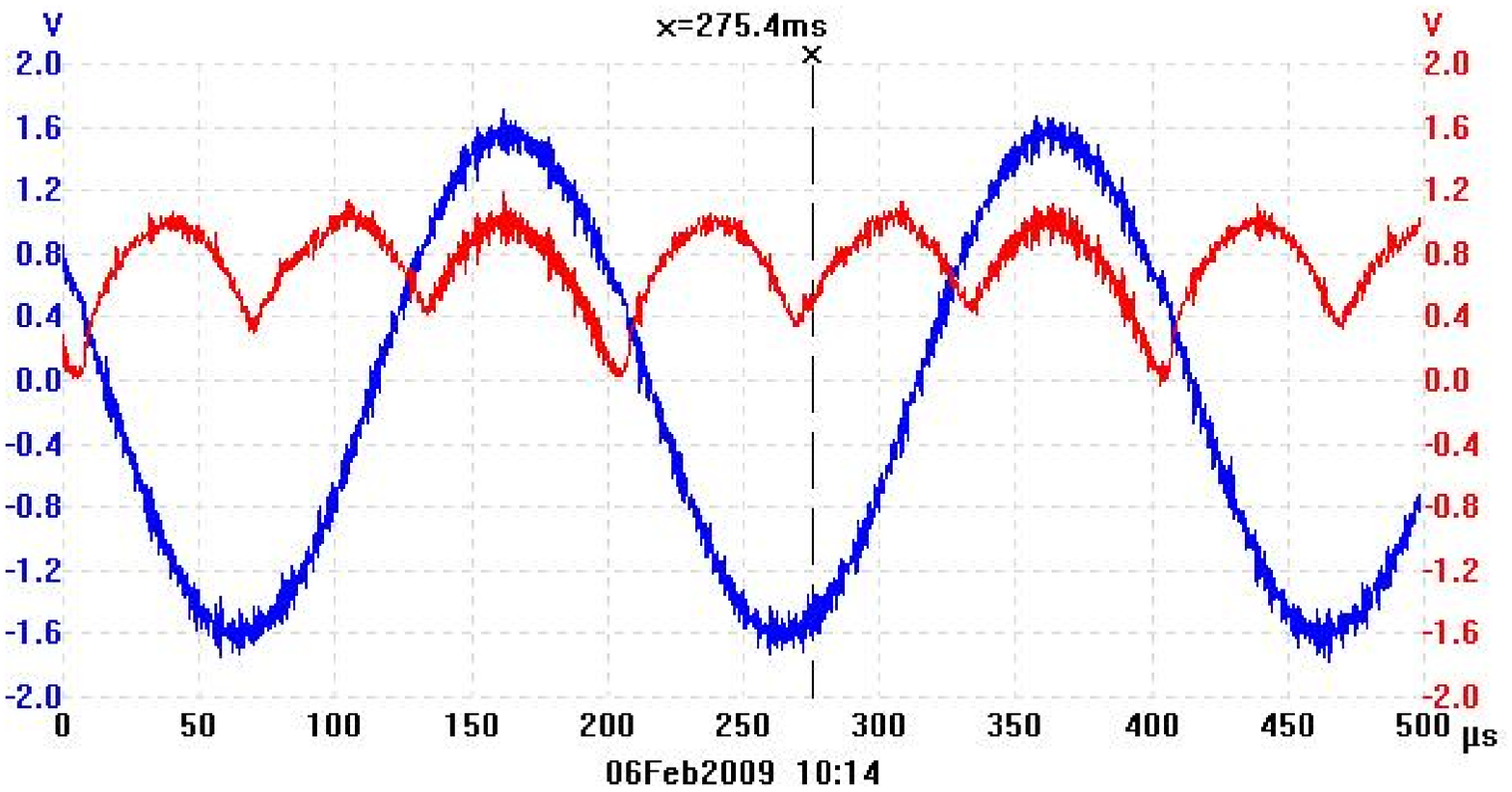}
\caption{A half-wave rectification circuit for three phase supply. The 
output waveform of this circuit shows that the ripples would be less than the 
case of half-wave rectification of a single phase supply.}
\end{center}
\end{figure}

Fig~4 shows the waveforms (phase A and C) as captured by Picoscope (Model
2202). Since the model only gives dual trace, only two waves can be shown
simultaneously. However, the data captured by Picoscope shows all three
phases simultaneously. Curve fitting these data points give the phase
difference between the various phases to be 1.95radians or ${\rm \sim
112^o}$. Also, the peak voltage (${\rm V_m}$) also works out to be 1.58v.

While the circuits achieves the purpose of mimicing a 3-phase supply,
experiments can be done to further understand applications of 3-phase
supply. For example, one can study the rectification and advantage of
generating DC from 3-phase supply. For this, all one needs is 3 diodes and a
load resistance (see fig~5). The output waveform is DC and also notice 
even without any filtering circuit, the output is continuous with
little ripple. This is because, as seen by the load, the input frequency is
three times that of a single phase supply. The ripples in a rectified output 
is inversely proportional to the frequency \cite{arun}, hence the
low ripples here can be understood. Also, this output gives visual idea and 
helps in easier understanding that each phase in a 3-phase supply is 
separated by ${\rm 120^o}$. In this rectifier circuit, only that diode
conducts, for which the phase connected to it has the highest instantaneous
potential with respect to the neutral. Fig~4 shows one phase to have highest
potential with repect to the other phases for ${\rm 120^o}$. Thus, in one
cycle (${\rm 2\pi}$) there would be three peaks of the output
DC wave. This is visiable in fig~5.

\section*{Conclusion}
A simple circuit has been proposed to demonstrate the behaviour of 3-phase
power distribution. The simple circuit discussed in this article cost Rs 
(Indian Rupee) 20/- to assemble (less than a dollar) and 
can be a very usual experiment in
schools and under-graduate laboratorys to learn more about 
3-phase supply and it's
application. 

\section*{Acknowledgments}

The financial support of U.G.C (India) in the form of Minor Research
Project No.F.6-1(25)/2007(MRP/SC/NRCB) is gratefully  
acknowledged. The authors would like to express their gratitude to the lab
technicians of
the Department of Physics and Electronics, S.G.T.B. Khalsa College, for the help
rendered in carrying out the experiment.

\end{document}